\documentstyle[aps]{revtex}

\begin{document}
\author{O. G. Balev}
\address{Departamento de F\'{\i}sica, Universidade Federal de S\~{a}o Carlos,%
\\
13565-905, S\~{a}o Carlos, S\~{a}o Paulo, Brazil \ and Institute of\\
Semiconductor Physics, National Academy of Sciences, 45 Pr. Nauky, Kiev\\
252650, Ukraine }
\author{Nelson Studart}
\address{Departamento de F\'{\i}sica, Universidade Federal de S\~{a}o Carlos,%
\\
13565-905, S\~{a}o Carlos, S\~{a}o Paulo, Brazil }
\title{Electron correlation effects in a wide channel from the $\nu=1$
quantum Hall edge states}
\maketitle

\begin{abstract}
The spatial behavior of Landau levels (LLs) for the $\nu =1$ quantum Hall
regime at the edge of a wide channel is studied in a self-consistent way by
using a generalized local density approximation proposed here. Both exchange
interaction and strong electron correlations, due to edge states, are taken
into account. They essentially modify the spatial behavior of the occupied
lowest spin-up LL in comparison with that of the lowest spin-down LL, which
is totally empty. The contrast in the spatial behavior can be attributed to
a different effective one-electron lateral confining potentials for the
spin-split LLs. Many-body effects on the spatially inhomogeneous
spin-splitting are calculated within the screened Hartree-Fock
approximation. It is shown that, far from the edges, the maximum activation
energy is dominated by the gap between the Fermi level and the bottom of the
spin-down LL, because the gap between the Fermi level and the spin-up LL is
much larger. In other words, the maximum activation energy in the bulk of
the channel corresponds to a highly asymmetric position of the Fermi level
within the gap between spin-down and spin-up LLs in the bulk. We have also
studied the renormalization of the edge-state group velocity due to electron
correlations. The results of the present theory are in line with those
suggested and reported by experiments on high quality samples.\newline
\end{abstract}

\section{INTRODUCTION}

Even though most previous theoretical works have developed a noninteracting
picture of the edge states in quantum Hall regime, the influence of
electron-electron interactions on the edge-state properties in a channel %
\cite{chklovskii92,dempsey93,gelfand94,brey93,muller92} and on the subband
structure of quantum wires\cite{kinaret90,susuki93,wrobel94},\cite{brey93}
has been the subject of intense study in the recent years. In Refs. \cite%
{chklovskii92,brey93,susuki93}, only the direct Hartree interaction was
taken into account. Nevertheless, edge states played no fundamental role in
many studies that develop or use theoretical pictures of the spin-splitting
Land\'{e} $g^{\ast }$ factor enhanced by the exchange interaction \cite%
{ando74,nicholas88,callin84,fogler95,xu95,leadley98,drichko99,brosig00}. On
the other hand, in Ref. \cite{balev97}, correlation effects due to edge
states on the effective $g^{\ast }$ factor for a quantum wire at the $\nu =1$
quantum Hall regime were considered, while for a wide channel only some
qualitative estimation was given. It was shown that correlations, due to the
screening at the edges, strongly suppress the exchange splitting and
smoothen the energy dispersion at the edges.\cite{balev97}

This paper provides a step further towards the understanding of the role of
electron-electron interactions in the $\nu =1$ quantum Hall effect regime,
in particular, the very important role of edge states, by extending the
approach of Ref. \cite{balev97}, which was based on the screened
Hartree-Fock approximation (SHFA) \cite{ando74,halperin68,ando82}. By using
the SHFA, we develop a generalized local density approximation (GLDA), which
is an essential improvement of the previous modified local density
approximation (MLDA) of Ref. \cite{balev97}. The main objective is to
determine the enhanced spin-splitting for the two-dimensional electron
system (2DES) in a wide channel and the position of the Fermi level within
the {\it exchange enhanced} gap. A realistic model for the edge regions of a
wide channel in a strong magnetic field ${\bf B}$ is solved
self-consistently when the lowest spin-polarized Landau level (LL)is
occupied, i.e., when $\nu =1$ in the inner part of the channel and the
formation of a dipolar strip\cite{chklovskii92} at the channel edges does
not occur. Moreover, we assume that a bare confining potential is rather
steep that prevents the flattening of edge states\cite%
{chklovskii92,dempsey93,brey93,susuki93} in the vicinity of the Fermi level.
The confining potential of the model is obtained in the Hartree
approximation as a self-consistent one-electron potential\cite%
{brey93,susuki93}, which in addition to a bare confining potential includes
the screening by the 2DES and pertinent electron-electron interactions. As
we assume that there is no flattening at the edge regions, the slope of the
confining potential, proportional to the group velocity $v_{g}^{H}$ in the
Hartree approximation, is finite. It is also assumed that the confining
potential, without many-body interactions, is smooth on the $\ell _{0}$
scale, where $\ell _{0}=(\hbar /m\omega _{c})^{1/2}$ with $\omega
_{c}=|e|B/mc$ the cyclotron frequency, and, hence, leads to a rather small $%
v_{g}^{H}$.

In this work we show that, if we go beyond the exchange interaction, the
spatial behavior of the LLs is strongly modified between the middle part of
the channel and the region near the edges due to electron correlations.
Furthermore, the position of the Fermi level in the gap at the inner part of
the channel is highly asymmetric. In the region, where the LLs are flat, the
Fermi level is much closer to the upper spin-split LL than to the occupied
lower spin-split LL. The most essential role played by correlations is
related to the screening by the edge states which in turn depends strongly
on their group velocity $v_{g}$. We notice that the exchange interaction
leads to an infinite (logarithmically divergent) $v_{g}$. Electron
correlations may restore a smooth dispersion of the single-particle energy,
on the $\ell _{0}$ scale, as a function of the oscillator center $%
y_{0}\approx k_{x}\ell _{0}^{2}$. Because, in typical experiments, the
condition of strong $B$ limit, $r_{0}=e^{2}/(\varepsilon \ell _{0}\hbar
\omega _{c})\ll 1$, is not satisfied, we point out that the proposed GLDA
gives and adequate self-consistent treatment to many-body effects in a {\em %
strong} $B$, when $r_{0}\alt1$. The paper is organized as follows. In Sec.
II, we describe a new microscopic nonperturbative approach for the
calculation of the screened Coulomb interaction for laterally nonhomogeneous
2DES in a wide channel in the strong magnetic field limit $r_{0}\ll 1$. In
Sec. III, we study exchange correlations effects for $\nu =1$ by using a
proposed GLDA to obtain the activation gap, the asymmetry of the Fermi level
position within the Fermi gap and investigate the renormalization of the LLs
due to exchange and correlations for the case more experimentally feasible
strong $B$, i.e. $r_{0}\alt1.$ In addition, we apply our theory to the
experiment described in Ref.\cite{nicholas88}. We summarize our results with
some remarks in Sec. IV.

\section{EXCHANGE-CORRELATION EFFECTS IN A WIDE CHANNEL}

We consider a strictly two-dimensional electron system confined in the $%
(x,y) $ plane to a wide channel of width $W$ and length $L_{x}=L$, in the
presence of a strong magnetic field $B$ along the $z$ axis. Taking the
vector potential ${\bf A}=-By\widehat{{\bf x}}$, we write the
single-particle Hamiltonian as $\widehat{h}^{0}=[(\widehat{p}_{x}+eBy/c)^{2}+%
\widehat{p}_{y}^{2}]/2m^{\ast }+V_{y}+g_{0}\mu _{B}\widehat{\sigma }_{z}B/2$%
, where the confining potential $V_{y}=0$ at the inner part of the channel
and $V_{y}=m\Omega ^{2}(y-y_{r})^{2}/2$, $y\geq y_{r}$, at the right edge,
with $\Omega \ll \omega _{c}$; $g_{0}$ being the bare Land\'{e} g-factor, $%
\mu _{B} $ the Bohr magneton, and $\hat{\sigma}_{z}$ $z$-component Pauli
matrix. The eigenvalues and eigenfunctions near the right edge of the
channel are well approximated by $\epsilon _{n,k_{x},\sigma }=(n+1/2)\hbar
\omega _{c}+m^{\ast }\Omega ^{2}(y_{0}-y_{r})^{2}\Theta
(y_{0}-y_{r})/2+\sigma g_{0}\mu _{B}B/2$ and $\psi _{nk_{x}\sigma }({\bf r}%
;\sigma _{1})=\left\langle {\bf r}|nk_{x}\right\rangle |\sigma >$, with $%
\left\langle {\bf r}|nk_{x}\right\rangle =\exp (ik_{x}x)\Psi
_{n}(y-y_{0}(k_{x}))/\sqrt{L}$ and spin function $|\sigma >=\psi _{\sigma
}(\sigma _{1})=\delta _{\sigma \sigma _{1}}$, $\sigma _{1}=\pm 1$; here $%
y_{0}\equiv y_{0}(k_{x})=\ell _{0}^{2}k_{x}$, $\Psi _{n}(y)$ is a harmonic
oscillator function, $\Theta (x)=1$ for $x>0$ and $\Theta (x)=0$ for $x<0$, $%
{\bf r}=\{x,y\}$. The edge of the ($n,\sigma $) LL is denoted by $%
y_{rn}^{(\sigma )}=y_{r}+\ell _{0}^{2}k_{e}^{n,\sigma }=\ell
_{0}^{2}k_{rn}^{(\sigma )}$, where $k_{rn}^{(\sigma )}=k_{r}+k_{e}^{n,\sigma
}$, and $W=2y_{r0}^{(1)}$ and the group velocity of the edge states $%
v_{gn}^{\sigma ,H}=\partial \epsilon _{n,k_{r}+k_{e}^{n,\sigma }}/\hbar
\partial k_{x}=\hbar \Omega ^{2}k_{e}^{n,\sigma }/m\omega _{c}^{2}$ with
wave vector $k_{e}^{n,\sigma }=(\omega _{c}/\hbar \Omega )\sqrt{2m^{\ast
}\Delta _{Fn}^{(\sigma )}}$, $\Delta _{Fn}^{(\sigma
)}=E_{F}^{H}-(n+1/2)\hbar \omega _{c}-\sigma g_{0}\mu _{B}B/2$, and $%
E_{F}^{H}$ is the Fermi energy in the Hartree approximation. We can also
write $v_{gn}^{\sigma ,H}=cE_{en}^{(\sigma )}/B$, where $E_{en}^{(\sigma
)}=\Omega \sqrt{2m^{\ast }\Delta _{Fn}^{(\sigma )}}/|e|$ is the electric
field associated with the confining potential $V_{y}$ at $y_{rn}^{(\sigma )}$%
. We also introduced the wave vector $k_{r}=y_{r}/\ell _{0}^{2}$. For
definiteness, we take the background dielectric constant $\varepsilon $ to
be spatially homogeneous. Because we will apply our theory for the case of
GaAs based samples, where $g_{0}=-0.44$, it is assumed in our study that $%
g_{0}<0$.

We begin by considering the strong magnetic field limit when $r_{0}\ll 1$
and only the lowest upper-spin level is occupied. It was shown in Ref. \cite%
{balev97}, that the exchange and correlation contributions to the
single-particle energy $E_{0,k_{x},1}=\epsilon _{0,k_{x},1}+\epsilon
_{0,k_{x},1}^{xc}$ in the SHFA can be written as

\begin{equation}
\epsilon _{0,k_{x},1}^{xc}=-\frac{1}{8\pi ^{3}}%
\int_{-k_{r0}^{(1)}}^{k_{r0}^{(1)}}dk_{x}^{^{\prime }}\int_{-\infty
}^{\infty }dq_{y}\int_{-\infty }^{\infty }dq_{y}^{^{\prime
}}V^{s}(k_{-},q_{y};q_{y}^{^{\prime }})
(0k_{x}|e^{iq_{y}y}|0k_{x}^{^{\prime}}) (0k_{x}^{^{\prime
}}|e^{iq_{y}^{^{\prime }}y}|0k_{x}),  \label{1}
\end{equation}
where $V^{s}(q_{x},q_{y};q_{y}^{^{\prime }})$ is the Fourier transform of
the screened Coulomb interaction which can be evaluated within the random
phase approximation (RPA), $k_{\pm}=k_x\pm k^{^{\prime}}_x$, and the matrix
element $(0k_{x}|\exp[iq_{y}y]|0k_{x}^{^{\prime}})=
\exp\{-[k_{-}^{2}+q_{y}^{2}-2iq_{y}k_{+}]\ell_{0}^{2}/4\}$. Notice that
close to the edges, if neglect by screening in $V^{s}(q_{x},q_{y};q_{y}^{^{%
\prime }})$, Eq. (\ref{1}) takes into account exchange in the first order of 
$r_{0}$, while in the inner part of the channel it is taken into account
practically exactly.

In order to evaluate the screened interaction, we use the self-consistent
field form of the random phase approximation (RPA).\cite{platzmann73} Let us
consider the static screened potential $\varphi (x-x_{0},y;y_{0})$, where
the argument $(x-x_{0})$ takes into account translational invariance in the $%
x$ direction, of an electron charge $e\delta (\vec{r}-\vec{r}_{0})$ located
at ($x_{0},y_{0}$). The one-electron Hamiltonian, in the presence of a
self-consistent potential $V^{s}(x-x_{0},y;y_{0})=e\varphi (x-x_{0},y;y_{0})$%
, is $\hat{H}=\hat{h}^{0}+V^{s}(x-x_{0},y;y_{0})$. The equation of motion
for the one-electron density matrix $\hat{\rho}$ is solved together with
Poisson's equation for the self-consistent potential. Following closely the
approach of Ref. \cite{balev99}, we obtain the integral equation for the
Fourier components of the induced charge density as

\begin{eqnarray}
&&\rho (q_{x},y;y_{0})=\frac{2e^{2}}{\varepsilon L}\sum_{n_{\alpha }=0}^{%
\bar{n}}\sum_{k_{x\alpha }}\sum_{\sigma _{\alpha }}F_{n_{\alpha },\sigma
_{\alpha }}^{(0)}(k_{x\alpha },q_{x})\ \Pi _{n_{\alpha }n_{\alpha
}}(y,k_{x\alpha },k_{x\alpha }-q_{x})  \nonumber \\
&&\times \int_{-\infty }^{\infty }d\tilde{y}\int_{-\infty }^{\infty
}dy^{\prime }\ \Pi _{n_{\alpha }n_{\alpha }}(\tilde{y},k_{x\alpha
},k_{x\alpha }-q_{x})K_{0}(|q_{x}||\tilde{y}-y^{\prime }|)\ [\rho
(q_{x},y^{\prime };y_{0})+e\delta (y^{\prime }-y_{0})],  \label{2}
\end{eqnarray}
where

\[
F_{n_{\alpha },\sigma _{\alpha }}^{(0)}(k_{x\alpha },q_{x})=\frac{%
f_{n_{\alpha },k_{x\alpha }-q_{x},\sigma _{\alpha }}-f_{n_{\alpha
},k_{x\alpha },\sigma _{\alpha }}}{\epsilon _{n_{\alpha },k_{x\alpha
}-q_{x},\sigma _{\alpha }}-\epsilon _{n_{\alpha },k_{x\alpha },\sigma
_{\alpha }}+i\hbar /\tau }, 
\]
and $\Pi _{n_{\alpha }n_{\beta }}(y,k_{x\alpha },k_{x\beta })=\Psi
_{n_{\alpha }}(y-y_{0}(k_{x\alpha }))\Psi _{n_{\beta }}(y-y_{0}(k_{x\beta
})) $. Here $f_{n_{\alpha },k_{x\alpha },\sigma _{\alpha }}$ is the
Fermi-Dirac function, $\bar{n}$ denotes the highest occupied LL and $%
K_{0}(x) $ is the modified Bessel function. To obtain Eq. (\ref{2}), we have
used\ the condition $q_{x}v_{g0}^{1,H}\ll \omega _{c}$, which should be well
satisfied for actual $q_{x}\alt\ell _{0}^{-1}$ due to the smoothness of the
confining potential.

For $\nu =1$ we have $\bar{n}=0$, $\sigma _{\alpha }=1$ and Eq. (\ref{2})
takes the form

\begin{eqnarray}
&&\rho (q_{x},y;y_{0})=\frac{e^{2}}{\pi \varepsilon }\int_{-\infty }^{\infty
}dk_{x\alpha }F_{0,1}^{(0)}(k_{x\alpha },q_{x})\ \Pi _{00}(y,k_{x\alpha
},k_{x\alpha }-q_{x})  \nonumber \\
&&\times \int_{-\infty }^{\infty }d\tilde{y}\int_{-\infty }^{\infty
}dy^{\prime }\ \Pi _{00}(\tilde{y},k_{x\alpha },k_{x\alpha
}-q_{x})K_{0}(|q_{x}||\tilde{y}-y^{\prime }|)\ [\rho (q_{x},y^{\prime
};y_{0})+e\delta (y^{\prime }-y_{0})].  \label{3}
\end{eqnarray}
Assuming that actual $|q_{x}|\ll k_{e}^{0,1}$ and considering only the right
edge region or the inner part of the channel, we can approximate the
numerator of the expression for $F_{0,1}^{(0)}(k_{x\alpha },q_{x})$ as $%
(f_{0,k_{x\alpha }-q_{x},1}-f_{0,k_{x\alpha },1})\approx -q_{x}(\partial
f_{0,k_{x\alpha },1}/\partial k_{x\alpha })=q_{x}\delta (k_{x\alpha
}-k_{r0}^{(1)})$. In addition, taking into account the smoothness of $%
\epsilon _{0,k_{x\alpha },1}$ at the edge (i.e., $|\epsilon _{0,k_{x\alpha
}-q_{x},1}-\epsilon _{0,k_{x\alpha },1}|/\epsilon _{0,k_{x\alpha },1}\ll 1$
for $|q_{x}|\alt\ell _{0}^{-1}$), it follows that $F_{0,1}^{(0)}(k_{x\alpha
},q_{x})\approx -\delta (k_{x\alpha }-k_{r0}^{(1)})/\hbar v_{g0}^{1,H}$.
Substituting the latter in Eq. (\ref{3}) and then integrating over $%
k_{x\alpha }$, we obtain

\begin{eqnarray}
&&\rho (q_{x},y;y_{0})=-r_{1}^{H}\;\Pi
_{00}(y,k_{r0}^{(1)},k_{r0}^{(1)}-q_{x})\int_{-\infty }^{\infty }d\tilde{y}%
\int_{-\infty }^{\infty }dy^{\prime }\   \nonumber \\
&&\times \Pi _{00}(\tilde{y},k_{r0}^{(1)},k_{r0}^{(1)}-q_{x})K_{0}(|q_{x}||%
\tilde{y}-y^{\prime }|)\ [\rho (q_{x},y^{\prime };y_{0})+e\delta (y^{\prime
}-y_{0})],  \label{4}
\end{eqnarray}
where $r_{1}^{H}=e^{2}/(\pi \hbar \varepsilon v_{g0}^{1,H})$ is a
characteristic dimensionless parameter for the system. Notice that for the
assumed symmetric wide channel, one can neglect the effect of the left edge
states on the right edge region or the inner part of the channel. The
solution of Eq. (\ref{4}) can be written as

\begin{equation}
\rho (q_{x},y;y_{0})=\rho (q_{x},y_{0})\;\Psi _{0}(y-y_{r0}^{(1)})\Psi
_{0}(y-y_{r0}^{(1)}+q_{x}\ell _{0}^{2}),  \label{5}
\end{equation}
where

\begin{eqnarray}
\rho (q_{x},y_{0}) &=&-e\ r_{1}^{H}[1+r_{1}^{H}M(0,q_{x})]^{-1}\   \nonumber
\\
&&\times \int_{-\infty }^{\infty }dx\Psi _{0}(x)\Psi _{0}(x+q_{x}\ell
_{0}^{2})K_{0}(|q_{x}||x-(y_{0}-y_{r0}^{(1)})|).  \label{6}
\end{eqnarray}
Here $M(0,q_{x})=\exp (-q_{x}^{2}\ell _{0}^{2}/4)K_{0}(q_{x}^{2}\ell
_{0}^{2}/4)/2$ is a special case of the matrix element

\begin{equation}
M(k_{x}-k_{r0},q_{x})=e^{-q_{x}^{2}\ell _{0}^{2}/2}\int_{0}^{\infty }dq_{y}%
\frac{e^{-q_{y}^{2}\ell _{0}^{2}/2}}{\sqrt{q_{x}^{2}+q_{y}^{2}}}\cos
[q_{y}(k_{x}-k_{r0})\ell _{0}^{2}],  \label{7}
\end{equation}
where $M(x,y)\equiv M(x,-y)\equiv M(-x,y)$.

Substituting Eq. (\ref{6}) into Eq. (\ref{5}), using the latter in the
Poisson's equation for the total electric potential $\varphi (q_{x},y;y_{0})$
induced by the total charge density $[\rho (q_{x},y;y_{0})+e\delta
(y-y_{0})] $, and Fourier transforming the result, we obtain the screened
Coulomb potential $V^{s}(q_{x},q_{y};q_{y}^{\prime })=e\varphi
(q_{x},q_{y};q_{y}^{\prime })$ as

\begin{eqnarray}
V^{s}(q_{x},q_{y};q_{y}^{\prime }) &=&\frac{2\pi e^{2}}{\varepsilon \sqrt{%
q_{x}^{2}+q_{y}^{2}}}{\LARGE \{}2\pi \delta (q_{y}+q_{y}^{\prime })-\frac{%
\pi r_{1}^{H}}{\sqrt{q_{x}^{2}+(q_{y}^{\prime })^{2}}}e^{-q_{x}^{2}\ell
_{0}^{2}/2}\   \nonumber \\
&&\times e^{-[q_{y}^{2}+(q_{y}^{\prime })^{2}]\ell _{0}^{2}/4}\;\frac{\exp
[-i(q_{y}+q_{y}^{\prime })(2k_{r0}^{(1)}-q_{x})\ell _{0}^{2}/2]}{%
1+r_{1}^{H}M(0,q_{x})}{\LARGE \}}.  \label{8}
\end{eqnarray}%
The first term in the curly brackets of Eq. (\ref{8}) is the bare Coulomb
interaction which leads to the exchange contribution. Substituting it in Eq.(%
\ref{1}), leads to\cite{balev97}

\begin{equation}
\epsilon _{0,k_{x},1}^{x}=-\frac{e^{2}}{2\pi \varepsilon \ell _{0}}\int_{%
\tilde{k}_{x}-\tilde{k}_{r0}^{(1)}}^{\tilde{k}_{x}+\tilde{k}%
_{r0}^{(1)}}dte^{-t^{2}/4}K_{0}(t^{2}/4),  \label{9}
\end{equation}%
where $\tilde{k}_{x,r0}=k_{x,r0}\ell _{0}.$ For $\tilde{k}_{r0}^{(1)}-|%
\tilde{k}_{x}|\gg 1,$ $\epsilon _{0,k_{x},1}^{x}\approx \epsilon
_{0}^{x}=-(\pi /2)^{1/2}(e^{2}/\varepsilon \ell _{0})$ and $\epsilon _{0,\pm
k_{r0}^{(1)},1}^{x}=\epsilon _{0}^{x}/2$, for $\tilde{k}_{x}=\pm \tilde{k}%
_{r0}^{(1)}$.

Now, substituting Eq. (\ref{8}) into Eq. (\ref{1}), we obtain, for the right
edge region or the inner part of channel, that

\begin{eqnarray}
\epsilon _{0,k_{x},1}^{xc} &=&-\frac{e^{2}}{\pi \varepsilon }\int_{-\infty
}^{k_{r0}^{(1)}}dk_{x}^{\prime }\frac{1}{1+r_{1}^{H}M(0,k_{x}-k_{x}^{\prime
})}\   \nonumber \\
&&\times \left\{ M(0,k_{x}-k_{x}^{\prime
})+r_{1}^{H}[M^{2}(0,k_{x}-k_{x}^{\prime
})-M^{2}(k_{x}-k_{r0}^{(1)},k_{x}-k_{x}^{\prime })]\right\} .  \label{10}
\end{eqnarray}
The first two positive terms in the curly brackets of Eq. (\ref{10}) lead to
the exchange contribution given by Eq. (\ref{9}). The third negative term is
the important contribution coming from electron correlations.

In order to estimate the correlations role, let us take $\epsilon
_{0,k_{x},1}^{xc}$ at the Fermi level, i.e., for $k_{x}=k_{r0}^{(1)}$. Then,
from Eq. (\ref{10}), it follows that

\begin{eqnarray}
\epsilon _{0,k_{r0}^{(1)},1}^{xc} &=&-\frac{e^{2}}{\pi \varepsilon \ell _{0}}%
\int_{0}^{\infty }dx[\exp (-x^{2}/4)K_{0}(x^{2}/4)]/2\   \nonumber \\
&&\times {\Large \{}1+r_{1}^{H}[\exp (-x^{2}/4)K_{0}(x^{2}/4)]/2{\Large \}}%
^{-1}.  \label{11}
\end{eqnarray}%
Notice that $r_{1}^{H}$ is typically a large parameter in GaAs ($\varepsilon
\approx 12.5$) based samples. Indeed, the characteristic velocity $e^{2}/\pi
\hbar \varepsilon \approx 5.6\times 10^{6}$ cm/s and due to its big value
usually can be considered much larger than $v_{g0}^{1,H}$. Then assuming
that $r_{1}^{H}\gg 1$ we obtain, from Eq. (\ref{11}), that $\epsilon
_{0,k_{r0}^{(1)},1}^{xc}\sim -\hbar v_{g0}^{1,H}/\ell _{0}$ which estimates
the many-body lowering of the Fermi level. Now, if we compare with the
Hartree-Fock approximation (HFA) result, we see that $\epsilon
_{0}^{x}/2\epsilon _{0,k_{r0}^{(1)},1}^{xc}\sim (\pi /2)^{3/2}r_{1}^{H}\gg 1$%
, which implies that correlations essentially contributes to diminish the
Fermi level lowering with respect to the bottom of the upper spin-split LL $%
(n=0,\sigma =-1)$.

\section{Correlation effects on the enhanced activation gap}

In the strong magnetic field limit, $r_{0}\ll 1$, the total single-particle
energy of the ($n=0,\sigma =1$) LL $E_{0,k_{x},1}=\epsilon
_{0,k_{x},1}+\epsilon _{0,k_{x},1}^{xc}$, where $\epsilon _{0,k_{x},1}^{xc}$
is given by Eq. (\ref{10}), can be written, in the right edge region or the
inner part of the channel, as

\begin{eqnarray}
E_{0,k_{x},1} &=&\frac{\hbar \omega _{c}}{2}-|g_{0}|\mu _{B}B/2+\frac{%
m^{\ast }\Omega ^{2}\ell _{0}^{4}}{2}(k_{x}-k_{r})^{2}\Theta (k_{x}-k_{r})-%
\frac{e^{2}}{\pi \varepsilon }\int_{-\infty }^{k_{r0}^{(1)}}dk_{x}^{\prime }%
\frac{1}{1+r_{1}^{H}M(0,k_{x}-k_{x}^{\prime })}\   \nonumber \\
&&\times \left\{ M(0,k_{x}-k_{x}^{\prime
})+r_{1}^{H}[M^{2}(0,k_{x}-k_{x}^{\prime
})-M^{2}(k_{x}-k_{r0}^{(1)},k_{x}-k_{x}^{\prime })]\right\} .  \label{12}
\end{eqnarray}
Notice that $\epsilon _{0,k_{r0}^{(1)},1}=E_{F}^{H}$ and only the ($%
n=0,\sigma =1$) LL is assumed to be occupied. Moreover, $E_{F}^{H}$ is the
quasi-Fermi level for this LL, when it appears above the bottom of the ($%
n=0,\sigma =-1$) LL. However, we demand that the quasi-Fermi level $E_{F}$
of the ($n=0,\sigma =1$) LL, renormalized by exchange, or by both exchange
and correlations, is the actual Fermi level, i.e., it is below the bottom of
($n=0,\sigma =-1$) LL. In addition, we assume that exchange correlation
effects do not change the Fermi wave vectors $k_{r0}^{(1)}$ and $k_{e}^{0,1}$%
. We point out that by taking the exchange interaction into account $E_{F}$
is already different from the $E_{F}^{H}$, since $|\epsilon _{0}^{ex}|/\hbar
\omega _{c}=\sqrt{\pi /2}r_{0}$, even for $r_{0}\ll 1$.

However, in actual experiments typically $r_{0}\sim 1.$ Now, we go beyond
the strong magnetic field limit for $E_{0,k_{x},1},$ given by Eq. (\ref{12}%
), by using $r_{1}=e^{2}/(\pi \hbar \varepsilon v_{g0}^{(1)})$ instead of $%
r_{1}^{H}$, and assuming that the approximation is still valid for $r_{0}\alt%
1$. Then Eq. (\ref{12}) converts into

\begin{eqnarray}
E_{0,k_{x},1} &=&\frac{\hbar \omega _{c}}{2}-|g_{0}|\mu _{B}B/2+\frac{%
m^{\ast }\Omega ^{2}\ell _{0}^{4}}{2}(k_{x}-k_{r})^{2}\Theta (k_{x}-k_{r})-%
\frac{e^{2}}{\pi \varepsilon }\int_{-\infty }^{k_{r0}^{(1)}-k_{x}}dx\frac{1}{%
1+r_{1}M(0,\sqrt{x^{2}+\delta ^{2}/\ell _{0}^{2}})}\   \nonumber \\
&&\times \left\{ M(0,\sqrt{x^{2}+\delta ^{2}/\ell _{0}^{2}})+r_{1}[M^{2}(0,%
\sqrt{x^{2}+\delta ^{2}/\ell _{0}^{2}})-M^{2}(k_{x}-k_{r0}^{(1)},\sqrt{%
x^{2}+\delta ^{2}/\ell _{0}^{2}})]\right\} ,  \label{13}
\end{eqnarray}
where the group velocity of edge states $v_{g0}^{(1)}=(\partial
E_{0,k_{x},1}/\hbar \partial k_{x})_{k_{x}=k_{r0}^{(1)}}$, renormalized by
exchange and correlations, which follows from the quadratic equation after
using Eq. (\ref{13}), is given by

\begin{equation}
v_{g0}^{(1)}=v_{g0}^{1,H}+\frac{e^{2}}{\pi \hbar \varepsilon }\frac{%
M(0,\delta /\ell _{0})}{1+r_{1}(v_{g0}^{(1)})\;M(0,\delta /\ell _{0})},
\label{14}
\end{equation}
which was calculated by considering that $[\partial M(k_{x}-k_{r0}^{(1)},%
\sqrt{x^{2}+\delta ^{2}/\ell _{0}^{2}})/\partial
k_{x}]_{k_{x}=k_{r0}^{(1)}}=0$. Equations (\ref{13}) and (\ref{14}) provide
the self-consistent scheme and are the basic equations of our GLDA. We have
introduced $\delta \ll 1$ in the second argument of $M(a,x),$ by changing $%
x=k_{x}^{\prime }-k_{x}$ to $\sqrt{x^{2}+\delta ^{2}/\ell _{0}^{2}}$, in
order to avoid very weak logarithmic divergence for $x\rightarrow 0$.
Indeed, $M(0,x)\approx \lbrack \ln (2\sqrt{2}/\ell _{0}|x|)-\gamma /2]$, for 
$x\rightarrow 0$, where $\gamma $ is the Euler constant. As one will see,
the results are very weakly dependent on $\delta $. This small parameter can
be estimated as $\delta \sim \max [\ell _{0}/d;\ell _{0}/v_{g0}^{(1)}\bar{%
\tau}]$, where $d$ is a typical distance of a remote screening region (or a
gate) and $\bar{\tau}$ is a typical lifetime at the edge states.

From Eq. (\ref{14}), we obtain that exists only one root

\begin{equation}
v_{g0}^{(1)}=\frac{v_{g0}^{1,H}}{2}\left[ 1+\sqrt{1+4r_{1}^{H}M(0,\delta
/\ell _{0})}\right] \approx \sqrt{\ln (2\sqrt{2}/\delta )}\sqrt{\frac{e^{2}}{%
\pi \hbar \varepsilon }v_{g0}^{1,H}},  \label{15}
\end{equation}
satisfying the physical requirement $v_{g0}^{(1)}>0$, which means that the
LL is below $E_{F}$ in the region between the middle and the right edge of
the channel $y_{r0}^{(1)}$. From Eq. (\ref{15}), we obtain that $%
v_{g0}^{(1)}/v_{g0}^{1,H}\approx \lbrack \ln (2\sqrt{2}/\delta
)r_{1}^{H}]^{1/2}\gg 1$. In addition, one see that $%
v_{g0}^{(1)}/v_{g0}^{1,H}\propto 1/\sqrt{v_{g0}^{1,H}}$.

The positive gap between the bottom of the upper spin-split LL and the Fermi
level of the interacting 2DES, $%
G(v_{g0}^{1,H})=E_{0,0,-1}-E_{0,k_{r0}^{(1)},1}$, is then given by

\begin{equation}
G=|g_{0}|\mu _{B}B-\frac{m^{\ast }\omega _{c}^{2}}{2\Omega ^{2}}%
(v_{g0}^{1,H})^{2}+\frac{e^{2}}{\pi \ell _{0}\varepsilon }\int_{0}^{\infty
}dt\frac{M(0,\sqrt{t^{2}+\delta ^{2}}/\ell _{0})}{1+R(v_{g0}^{1,H})M(0,\sqrt{%
t^{2}+\delta ^{2}}/\ell _{0})},  \label{16}
\end{equation}%
where $R(v_{g0}^{1,H})$ is the function which appears, from $%
r_{1}(v_{g0}^{(1)})$, after taking $v_{g0}^{(1)}$ in terms of $v_{g0}^{1,H}$
according to Eq. (\ref{15}) and $M(0,t/\ell _{0})=\exp
(-t^{2}/4)K_{0}(t^{2}/4)/2$. To obtain Eq. (\ref{16}), we put Eq. (\ref{13})
and the expression $k_{e}^{0,1}=(m^{\ast }\omega _{c}^{2}/\hbar \Omega
^{2})v_{g0}^{1,H}$ in the term $\propto (k_{r0}^{(1)}-k_{r})^{2}$. In the
inner part of the channel, the total gap between the lowest spin-split LLs
is $G_{-1,1}=(E_{0,0,-1}-E_{0,0,1})\approx |g_{0}|\mu _{B}B+\sqrt{\pi /2}%
(e^{2}/\varepsilon \ell _{0})$, where we have neglected many-body
contributions due to the weak ``bulk'' screening of the fully occupied LL as
previously discussed.\cite{balev97} However, while this ``bulk'' screening
effect is rather weak at the edge region in comparison with the edge-state
screening effect, it should be more significant in the inner part of the
channel far from the edge. Taking into account this effect, we obtain $%
G_{-1,1}=|g_{0}|\mu _{B}B+\sqrt{\pi /2}\cdot k_{bu}(r_{0})(e^{2}/\varepsilon
\ell _{0})$, where the numerical values of $k_{bu}(r_{0})$ smoothly vary
from $0.79$ to $0.63$ as $r_{0}$ goes from $0.6$ to $1.4.$ Then $%
(G_{-1,1}-|g_{0}|\mu _{B}B)/\hbar \omega _{c}=\sqrt{\pi /2}%
r_{0}k_{bu}(r_{0}) $ and values of $k_{bu}(r_{0})$ tend rather fast to $1$
as $r_{0}$ goes below $1$.

In the spirit of the MLDA\cite{balev97}, which has some similarity with the
local-density approximation (LDA) \cite{ando82,dempsey93b}, we claim that
the energy dispersion relation, given by Eqs. (\ref{13}) - (\ref{15}), comes
from the solution of the single-particle Schr\"{o}dinger equation (for $%
\sigma =1$) with the Hamiltonian $\hat{h}=\hat{h}^{0}+V_{xc}(y)$, where the
self-consistent exchange-correlation potential is given as

\begin{equation}
V_{xc}(y)=E_{0,y/\ell _{0}^{2},1}-(\frac{\hbar \omega _{c}}{2}-|g_{0}|\mu
_{B}B/2+V_{y}),  \label{17}
\end{equation}%
with $E_{0,x,1}$ being determined by Eqs. (\ref{13}) and (\ref{15}). The
validity of GLDA for $r_{0}\alt1$ is well justified if $v_{g0}^{(1)}/\ell
_{0}\ll \omega _{c}$. As a consequence, the eigenenergy $E_{0,x,1}$ as a
function of $x$ is smooth on the scale of $\ell _{0}$ for any occupied state
and the eigenfunction of any actual state can be again approximated by the
unperturbed $\psi _{nk_{x}\sigma }({\bf r};\sigma _{1})$. Using this, we
arrive to Eqs. (\ref{13}) and (\ref{15}), which reduces to Eq. (\ref{12}),
for $r_{0}\ll 1$ and $4r_{1}^{H}\ln (2\sqrt{2}/\delta )\ll 1$. However, as
stated before, the latter condition cannot be satisfied in GaAs based
samples. Now, assuming that $V_{xc}(y)$ is smooth on the scale of $\ell _{0}$
we find, neglecting small corrections, that the corresponding energy
dispersion is given back by Eqs. (\ref{13}) and (\ref{15}) for ($n=0,\sigma
=1$) LL, which confirms the successful self-consistent scheme of GLDA.
However, in contrast with the LDA, our $V_{xc}(y)$ depends essentially on
the slope at the edge $y_{r0}^{(1)}$ of the channel $\{d[V_{y}+V_{xc}(y)]/dy%
\}_{y=y_{r0}^{(1)}}\propto v_{g0}^{(1)}$, which can be quite different for
almost the same density profile of the 2DES. Moreover, this effect is
essential for some regions where $|y-y_{r0}^{(1)}|/\ell _{0}\gg 1$. Hence,
the GLDA incorporates nonlocal features, as the MLDA, but in contrast with
the LDA. The {\it effective one-electron confining potential} $%
V_{T}(y)=V_{y}+V_{xc}(y),$ for the ($n=0,\sigma =1$) LL, within GLDA is
determined, for $r_{0}\alt1,$ by Eqs. (\ref{13}), (\ref{15}), and (\ref{17})
and leads to strong modifications in the energy spectrum and activation gap,
while keeps the 2DES density profile practically constant, in comparison,
e.g., with the Hartree approximation result. We point out that the assumed
smoothness of the total confining potential for ($n=0,\sigma =1$) LL implies
that $v_{g0}^{(1)},$ given by Eq. (\ref{15}), should satisfy the condition $%
v_{g0}^{(1)}/\ell _{0}\ll \omega _{c}$. This condition can be rewritten as $%
[(r_{0}/\pi )(v_{g0}^{1,H}/\omega _{c}\ell _{0})\ln (2\sqrt{2}/\delta
)]^{1/2}\ll 1$, Then it is satisfied for typical $r_{0}\leq 1$, $\delta $
and $v_{g0}^{1,H}/\omega _{c}\ell _{0}\ll 1$.

Now, we define the dimensionless activation gap $%
G_{a}(v_{g0}^{1,H})=G/(|g_{0}|\mu _{B}B/2)$ due to exchange correlation
effects. In the absence of many-body interactions, the maximum value of $%
G_{a}=1$. So the activation gap is enhanced when $G_{a}>1$. The asymmetry of
the Fermi level position within the Fermi gap in the interior part of the
channel can be characterized by another dimensionless function $\delta
G(v_{g0}^{1,H})=(\bar{G}_{-1,1}-G_{a})/G_{a}$, where $\bar{G}%
_{-1,1}=G_{-1,1}/(|g_{0}|\mu _{B}B/2)$. Notice that, when $%
v_{g0}^{1,H}\rightarrow 0$, $E_{F}^{H}$ tends, from the above side, to the
bottom of the ($n=0,\sigma =1$) LL, in the absence of many-body interactions.

In Fig. 1, using Eq. (\ref{16}), $G_{a}$ is depicted as a function of $\bar{v%
}_{g0}^{1,H}=v_{g0}^{1,H}/(\hbar \omega _{c}/m^{\ast })^{1/2}$ for $\delta
=10^{-3}$, such that he condition of smoothness of the confining potential
is $\bar{v}_{g0}^{1,H}\ll 1$. The solid and dashed curves correspond to $%
\omega _{c}/\Omega =20$ and $10$, respectively. Furthermore, the solid
(dashed) curves correspond, from top to bottom, to $r_{0}=1.4,1.0$ and $0.6$%
, respectively. The maxima of $G_{a}$ for the solid curves, from top to
bottom, are $30.42$ (for $\bar{v}_{g0}^{1,H}\equiv \bar{v}%
_{g0}^{H,max}=0.0138$ and $k_{e}^{0,1}\ell _{0}=5.52$); $23.57$ (for $\bar{v}%
_{g0}^{H,max}=0.0119$ and $k_{e}^{0,1}\ell _{0}=4.7$); and $16.16$ (for $%
\bar{v}_{g0}^{H,max}=0.0095$ and $k_{e}^{0,1}\ell _{0}=3.80)$. For the
dashed curves, the maxima are, from top to bottom, $38.18$ (for $\bar{v}%
_{g0}^{H,max}=0.0297$ and $k_{e}^{0,1}\ell _{0}=2.97$); $29.30$ (for $\bar{v}%
_{g0}^{H,max}=0.0254$ and $k_{e}^{0,1}\ell _{0}=2.54$); and $19.75$ (for $%
\bar{v}_{g0}^{H,max}=0.0201$ and $k_{e}^{0,1}\ell _{0}=2.01$). Notice that
the maxima occur at values $\bar{v}_{g0}^{H,max}$ in which all assumed
conditions are well satisfied. When exchange and correlations are neglected, 
$E_{F}^{H}$ is a quasi-Fermi level at the maxima, because it is located
above the bottom of the ($n=0,\sigma =-1$) LL. Indeed, $E_{F}^{H}$ touches
the bottom of ($n=0,\sigma =-1$) LL at $\bar{v}_{g0}^{1,H}=0.0086$ and $%
k_{e}^{0,1}\ell _{0}=3.43$ for solid curve parameters and at $\bar{v}%
_{g0}^{1,H}=0.0172$ and $k_{e}^{0,1}\ell _{0}=1.72$ for the dashed curves,
if many-body contributions are neglected. In the present study, parameters
of GaAs-based sample are used, in particular, $\varepsilon \approx 12.5$ and 
$m^{\ast }=0.067m_{0}$.

In Fig. 2, we present $\delta G$ as a function of $\bar{v}_{g0}^{1,H}$ for
the same parameters used in Fig. 1. In the same way, the solid and dashed
curves correspond to $\omega _{c}/\Omega =20$ and $10$, respectively. At the
right side of the figure, the solid (dashed) curves correspond, from top to
bottom, to $r_{0}=0.6,$ $1.0$ and $1.4$, respectively. Notice that a maximum
value of $G_{a}$ and a minimum of $\delta G$ are at the same $\bar{v}%
_{g0}^{H,max}$. The minima $\delta G=4.11,$ $4.13$ and $4.00$ are for solid
curves while $\delta G=3.18,$ $3.13$ and $2.98$ for dashed lines (from top
to bottom at the right side). The results, shown in Figs. 1 and 2,
demonstrate that the position of the Fermi level, within the gap in the
interior part of the channel, is highly asymmetric and this will be even
stronger, if the effect of the ``bulk'' screening in $G_{-1,1}$ should be
neglected. So strong electron correlations of the edge states lead to a
highly asymmetric position of the Fermi level within the gap defined by the
two lowest spin-split LLs.

The results given in the solid curves of Figs. 1 and 2, for $r_{0}=1.4$,
correspond to the experiments of Ref. \cite{nicholas88}, with electron
density $n_{s}\approx 8.1\times 10^{10}$ cm$^{-2}$, in which a factor of
enhancement of the activation gap of the order of $15$ was observed. Indeed,
for the $\nu =1$ QHE regime, $B=2\pi \hbar cn_{s}/|e|\approx 3.34$ T gives $%
r_{0}\approx 1.4$ and $\omega _{c}\approx 8.76\times 10^{12}$ s$^{-1}$. In
our theory, $\Omega $ or $\omega _{c}/\Omega $ is an undetermined parameter
related to the parabolic confinement at the edges. It was estimated in Ref.%
\cite{muller92}, for $\nu =1$ and a sample with $n_{s}\approx 1.9\times
10^{11}$ cm$^{-2},$ as $\Omega \approx 7.8\times 10^{11}\;$s$^{-1}$.
However, Ref.\cite{wrobel94} gives a smaller value $\Omega \approx
4.16\times 10^{11}\;$s$^{-1}$, and here we implicitly assume that $\omega
_{c}/\Omega =20$ should be a realistic estimate for the sample of Ref.\cite%
{nicholas88}.

Now we present, in Fig. 3, $G_{a}$ for $r_{0}=1.4$, using Eq. (\ref{16}).
Solid and dashed curves correspond to $\delta =10^{-3}$ and $\delta =10^{-2}$%
, respectively. The solid (dashed) curves correspond, from top to bottom, to 
$\omega _{c}/\Omega =10,$ $20,$ $40,$ $60$ and $80$, respectively. One can
see that the maximum of $G_{a}$ for $\omega _{c}/\Omega =20$ is in
reasonable agreement with experiment.\cite{nicholas88} Observe that for $%
\omega _{c}/\Omega =60$ and $80,$ our result is very close to the observed
activation gap. We believe that the measured smaller value of the gap should
be related with effects of weak long-range inhomogeneities on the confining
potential $V_{y}$ at the inner part of the channel, which are absent in our
model. It is also seen that $G_{a}$ is weakly dependent on the small cutting
parameter $\delta $.

We show, in Fig. 4, numerical results, within GLDA, for LLs spectra $%
E_{0,k_{x},\pm 1}$ as function of $k_{x}$, where $k_{x}$ is measured as $%
(k_{x}-k_{r})\ell _{0}$, and $\tilde{k}_{x}=k_{x}\ell _{0}$ gives the
dimensionless oscillator center. We take $r_{0}=1.0$, $\omega _{c}/\Omega
=20 $, $\delta =10^{-3}$, $k_{e}^{0,1}\ell _{0}=(k_{r0}^{(1)}-k_{r})\ell
_{0}=4.76$, and $\bar{v}_{g0}^{1,H}=0.0119$, which are the parameters
corresponding to the middle solid curve of Fig. 1 at the maximum of $G_{a}.$
There, $\bar{v}_{g0}^{(1)}=v_{g0}^{(1)}/\sqrt{\hbar \omega _{c}/m^{\ast }}%
\approx 0.176\ll 1$ satisfies the ``smoothness'' requirement to
applicability of the GLDA. The bottom solid curve represents $E_{0,k_{x},1}$%
, from Eqs. (\ref{13}), (\ref{15}), and the dotted line gives the exact
position of $E_{F}$, when both exchange and correlations effects are taken
into account. The top solid curve is the spectrum of the upper spin-split LL 
$E_{0,k_{x},-1}$, and, for a sake of comparison, the close dashed-dotted
curve is $\epsilon _{0,k_{x},1}=[\hbar \omega _{c}+m^{\ast }\Omega
^{2}(y_{0}-y_{r})^{2}\Theta (y_{0}-y_{r})-\left| g_{0}\right| \mu _{B}B]/2$,
i.e., the spectrum of the lower spin-split LL without many-body
interactions. Finally, the dashed curve is $E_{0,k_{x},1}$ obtained within
the HFA where correlations effects are totally neglected. The horizontal
dashed line shows the position of $E_{F}$ within the HFA.

Similar curves to those in Fig. 4 are depicted in Fig. 5, but with the
parameters pertinent to experiments of Ref. \cite{nicholas88}. In Fig. 5,
the following parameters are used: $r_{0}=1.4$, $\omega _{c}/\Omega =20$, $%
\delta =10^{-3}$, $k_{e}^{0,1}\ell _{0}=5.52$, $\bar{v}_{g0}^{1,H}=0.0138$,
which are used to plot the top solid curve in Fig. 1 at its maximum. Notice
here $\bar{v}_{g0}^{(1)}\approx 0.224$ is small, which satisfies the
condition of ``smoothness'' for applicability of the GLDA. The small effect
coming from virtual interlevel transitions was neglected in Figs. 4 and 5 \
However, the most important contribution of such a small effect, related
with the 2DES screening in the middle of the channel for $\nu =1$ is taken
into account in Figs. 1-3.

\section{CONCLUDING REMARKS}

The self-consistent treatment developed here, for 2DES in the $\nu =1$ QHE
regime, shows that edge-state correlations drastically modify the ($%
n=0,\sigma =1$) LL spectrum in a wide region (with width $\gg \ell _{0}$)
nearby the channel edge, i.e. the effect of correlations induced by
edge-states is essentially nonlocal. Moreover, we have shown that the
position of the Fermi level $E_{F}$ is highly asymmetric within the gap
defined by the ($n=0,\sigma =1$) and ($n=0,\sigma =-1$) LLs in the interior,
or ``bulk'' part of the channel due to such correlations. The activation gap
is much smaller than the Fermi gap. In addition, we have obtained analytical
expressions for the activation gap and for the edge-state group velocity $%
v_{g0}^{(1)},$ renormalized by exchange and correlations effects.

We have obtained that $v_{g0}^{(1)}/v_{g}^{x}\approx \sqrt{\pi \bar{v}%
_{g0}^{1,H}/[r_{0}\ln (2\sqrt{2}/\delta )]}\ll 1$, where $%
v_{g}^{x}=(e^{2}/2\pi \hbar \varepsilon )K_{0}(\delta ^{2}/4)$ is the edge
group velocity in the HFA, $\bar{v}_{g}=v_{g}/\sqrt{\hbar \omega
_{c}/m^{\ast }}$ is the normalized group velocity, such that $\bar{v}_{g}\ll
1$ is the smoothness requirement for the pertinent confining potential, and $%
\delta $ is a small parameter. In particular, for the experimental
parameters of Ref. \cite{nicholas88}, used in Fig. 5, we calculate $%
v_{g0}^{(1)}/v_{g}^{x}\approx 0.06$. So while $\bar{v}_{g0}^{(1)}\approx
0.22\ll 1,$ the value of $\bar{v}_{g}^{x}\approx 3.60\gg 1$. Then, in this
case, edge-state correlations have decreased the group velocity by a factor
of $16$, as we can see from the slopes of pertinent solid and dashed curves
in Fig. 5, such that effective one-electron confining potential $V_{T}(y)$
is smooth.

About the role of correlations in the position of the Fermi level within the
gap between spin-down and spin-up LLs in the bulk, we point out that the
exchange interaction (HFA) leads the Fermi level to a position slightly
above the middle of the large exchange enhanced gap, tending to it only when 
$v_{g0}^{1,H}\rightarrow 0$. Because the correlations strongly reduce the
exchange contribution only nearby the edges, the physical consequence is
that the position of the Fermi level within the $\nu =1$ gap, should be
strongly asymmetric when both exchange and correlations are taken into
account. Indeed, in the actual situation, $v_{g0}^{1,H}/\sqrt{\hbar \omega
_{c}/m^{\ast }}<<1,$ this asymmetry comes from the fact that the many-body
contribution (the third term of Eq. (\ref{16})) goes $\propto v_{g0}^{1,H}$
while the Hartree term (the second of Eq. (\ref{16})) behaves like $%
(v_{g0}^{1,H})^{2}$. Only for relatively large $v_{g0}^{1,H}$, the Hartree
term becomes dominant because the many-body contribution tends to be
constant.

Even though we have used a simple analytical expression for the confining
potential, obtained in the Hartree approximation as the sum of the bare
confining potential and the Hartree potential, it often reproduces quite
well numerical results of the confining potential of a real wide channel
calculated in the Hartree approximation. Moreover, the above treatment can
be easily extended on the confining potential, obtained within Hartree
approximation, of a different form, if it is smooth on the scale of $\ell
_{0}$. Our results were obtained by developing a generalized local density
approximation (GLDA) in which the single-particle confining potential $%
V_{xc}(y)$ incorporates exchange and nonlocal correlation effects.

\acknowledgements

This work was supported in part by Funda\c{c}\~{a}o de Amparo \`{a} Pesquisa
do Estado de S\~{a}o Paulo (FAPESP). The authors are grateful to the
Conselho Nacional de Desenvolvimento Cient\'{\i}fico e Tecnol\'{o}gico
(CNPq) for research fellowships.

\bigskip

Fig. 1 Many-body enhancement of the activation gap $G_{a}=G/(\left|
g_{0}\right| \mu _{B}B/2)$ as a function of the group velocity $v_{g0}^{1,H}$%
, in the Hartree approximation, calculated from Eqs. (\ref{15}) and (\ref{16}%
). Solid (for $\omega _{c}/\Omega =20$) and dashed (for $\omega _{c}/\Omega
=10$) curves correspond to $r_{0}=1.4,\ 1.0,$ and $0.6,$ from top to bottom;
with $\delta =10^{-3}$. In the absence of many-body interactions, the
maximum of $G_{a}=1.$

\smallskip

Fig. 2 Fractional difference $\delta G=(G_{-1,1}-G)/G$, where $%
G_{-1,1}=E_{0,0,-1}-E_{0,0,1}$, as a function of $v_{g0}^{1,H}$, showing the
asymmetry of the Fermi level position within the Fermi gap, in the inner
part of the channel. Bulk screening effect is include in the calculation of
the total value of the Fermi gap $G_{-1,1}.$ The parameters and notations of
the curves are the same as in Fig. 1, but the solid and dashed curves here
correspond, counting at their right side from top to bottom, to $%
r_{0}=0.6,1.0,1.4$. Notice that the minimum of $\delta G=1$, when many-body
interactions are neglected.

\smallskip

Fig. 3 Activation gap $G_{a}$ as a function of $v_{g0}^{1,H}$ for $r_{0}=1.4$%
, corresponding to the experiment of Ref.\cite{nicholas88}. Curves are
shown, from top to bottom, for $\omega _{c}/\Omega =10,$ $20,$ $40,$ $60,$
and $80.$ The solid and dashed lines are given for $\delta =10^{-3}$ and $%
10^{-2}$, respectively. It is seen that $G_{a}$ is weakly dependent on the
small parameter $\delta $.

\smallskip

Fig. 4 Energy spectra as a function of $k_{x}$ for parameters $r_{0}=1.0$, $%
\omega _{c}/\Omega =20$, $\delta =10^{-3}$, $k_{e}^{0,1}\ell
_{0}=(k_{r0}^{(1)}-k_{r})\ell _{0}=4,76,$ and $v_{g0}^{1,H}=0.0119.$ These
values correspond to the maximum of $G_{a}$ shown by the middle solid curve
in Fig. 1. The solid curve at the bottom is $E_{0,k_{x},1}$, evaluated using
Eqs. (\ref{13}) and (\ref{15}), which can be compared with $E_{0,k_{x},1}$
in the Hartree-Fock approximation (no correlations involved) represented by
the dashed curve, and $\epsilon _{0,k_{x},1},$ the spectrum of the lower
spin-split LL without many-body interactions (dash-dotted curve). The dotted
line is the exact position of the Fermi level $E_{F}$, when
exchange-correlation effects are taken into account while the dashed line is
the position of the Fermi level in the Hartree-Fock approximation. The top
solid curve represents the spectrum of the upper-split LL, $E_{0,k_{x},-1}$.

\smallskip

Fig. 5 Same as in Fig. 4 with the assumed experimental parameters for the
sample of Ref. \cite{nicholas88}. Here $r_{0}=1.4$, $\omega _{c}/\Omega =20$%
, $\delta =10^{-3}$, $k_{e}^{0,1}\ell _{0}=5.52$, $\bar{v}_{g0}^{1,H}=0.0138$%
. These values correspond to those of the top solid curve in Fig. 1 at its
maximum.

\end{document}